\documentclass[runningheads]{llncs}

\usepackage[T1]{fontenc}
\usepackage{graphicx}
\usepackage{booktabs}
\usepackage{mathrsfs}
\usepackage{amsmath}
\usepackage{xcolor}
\usepackage{wrapfig}
\usepackage{float}
\usepackage[normalem]{ulem}
\usepackage[textsize=tiny]{todonotes}

\begin{document}

\title{ Neutrino Reconstruction in TRIDENT Based on Graph Neural Network}

\author{Cen Mo \inst{1}\thanks{Corresponding author. mo\_cen@sjtu.edu.cn}, Fuyudi Zhang \inst{2}, Liang Li\thanks{Corresponding author. liangliphy@sjtu.edu.cn} \inst{1}}

\institute{School of Physics and Astronomy, Shanghai Jiao Tong University
\and Tsung-Dao Lee Institute, Shanghai Jiao Tong University}

\maketitle

\begin{abstract}
TRopIcal DEep-sea Neutrino Telescope (TRIDENT) is a
next-generation neutrino telescope to be located in the South China Sea.
With a large detector volume and the use of advanced hybrid digital
optical modules (hDOMs), TRIDENT aims to discover multiple astrophysical neutrino sources and probe all-flavor neutrino physics. The reconstruction resolution of primary neutrinos is on the critical path to these
scientific goals. We have developed a novel reconstruction method based on graph neural network
(GNN) for TRIDENT. In this paper, we present the reconstruction performance of the GNN-based approach on both track- and shower-like
neutrino events in TRIDENT.
\end{abstract}


\section{Introduction}
In 2013, the first detection of astrophysical neutrinos was reported \cite{doi:10.1126/science.1242856}. 
Unlike cosmic rays, high-energy neutrinos remain unaffected by galactic magnetic fields, preserving their trajectory and pointing directly back to their sources. This makes them ideal instruments for investigating the origins of high-energy cosmic rays.

The deep inelastic scattering (DIS) between high-energy neutrinos and nucleons in water is employed to detect astrophysical neutrinos. 
When $\nu_\mu$ charged-current (CC) interactions occur, high-energy muons are generated and produce a kilometer-long track-like event topology. 
The track-like events are important in neutrino point-source searches, such as TXS 0506 \cite{txs} and NGC 1068 \cite{ngc1068}, due to their sub-degree level angular resolution.
On the other hand, $\nu_e$ CC interactions and neutral-current (NC) interactions produce a cascade of secondary particles at the DIS vertex, resulting in the deposition of neutrino energy in a localized region and forming a shower-like topology. 
Despite their poor angular resolution, the distinctive event topology of shower-like events makes them easily distinguishable from atmospheric-neutrino background. As such, shower-like events play a critical role in the search for extended neutrino sources.
For $\nu_\tau$ CC interactions, a tau lepton is generated along with a hadronic cascade.
The tau lepton travels some distance before decaying into a hadronic or electromagnetic cascade.
If the $\nu_\tau$ is sufficiently energetic, the two cascades resulting from the tau lepton's decay will be spatially separated, giving rise to a characteristic double cascade topology signature.
In the case of lower energy $\nu_\tau$ events, the identification of such events can be based on the presence of a double pulse in the readout waveform \cite{wille2019astrophysical}.

TRIDENT is a next-generation neutrino detector aiming to  identify astrophysical neutrino sources with high precision. 
This telescope design incorporates hybrid digital modules (hDOMs) comprising multiple Photomultiplier Tubes (PMTs) and Silicon Photomultipliers (SiPMs). 
To achieve comprehensive neutrino detection capabilities, these hDOMs are strategically planned for deployment across a vast cubic kilometer region deep in the deep waters of the South China Sea.

To reconstruct the direction and energy of incoming neutrinos using information from Cherenkov photons, both machine learning-based and likelihood-based reconstruction methods have been widely used in neutrino telescopes. 
In IceCube, convolutional neural networks (CNNs) ~\cite{abbasi_convolutional_2021}\cite{yu_trigger-level_2023} and GNNs~\cite{abbasi_graph_2022} have been assessed for their efficiency.
KM3NeT employs likelihood methods for both $\nu_e$ and $\nu_\mu$ in the reconstruction of direction and energy~\cite{km3netMLE}.
3D CNNs are also implemented in KM3NeT/ORCA~\cite{aiello_event_2020}. 
The likelihood method has relatively high reconstruction resolution but there is still room for improvement, especially in the case of $\nu_e$ events. The CNN approach faces challenges in handling sparse signals in TRIDENT which has a large detector volume. 

In this study, we propose a novel reconstruction method based on GNN.
We simulate $\nu_e$ CC and $\nu_\mu$ CC events utilizing the preliminary full detector configuration of TRIDENT. Subsequently, a GNN architecture is designed and employed to facilitate the precise reconstruction of direction for the neutrino events.

\section{Event Simulation}
The comprehensive Monte Carlo simulations of neutrino events are executed in two steps.

In the initial step, the DIS processes are simulated in the CORSIKA8 framework \cite{huege2022corsika}.
To represent the TRIDENT detector region, a cylindrical volume is constructed, with a radius of 2500 meters and a height of 1000 meters, positioned at a depth of 2900 meters below sea level.
The PYTHIA8 program \cite{bierlich2022comprehensive} is employed in CORSIKA8 to simulate the DIS processes.
By employing different rules for $\nu_e$ and $\nu_\mu$ neutrinos, accounting for their distinctive characteristics, the vertices are sampled accordingly.
Given that the typical size of hadronic cascades is less than 50 meters, to ensure an adequate number of Cherenkov photons for each event, the vertices of $\nu_e$ CC interactions are uniformly sampled within the detector region.
Conversely, high-energy muons exhibit significant travel distances in sea water. As a result, the DIS vertices of $\nu_\mu$ interactions are sampled over a larger region, the extent of which is contingent upon the energy of the muon involved.
Particles decay and propagate through water until they reach the detector region. 
Subsequently, the interactions of these particles and the response of detectors inside the telescope are further simulated using another dedicated program.

\begin{figure}[!htb]
    \centering\includegraphics[width=0.4\linewidth]{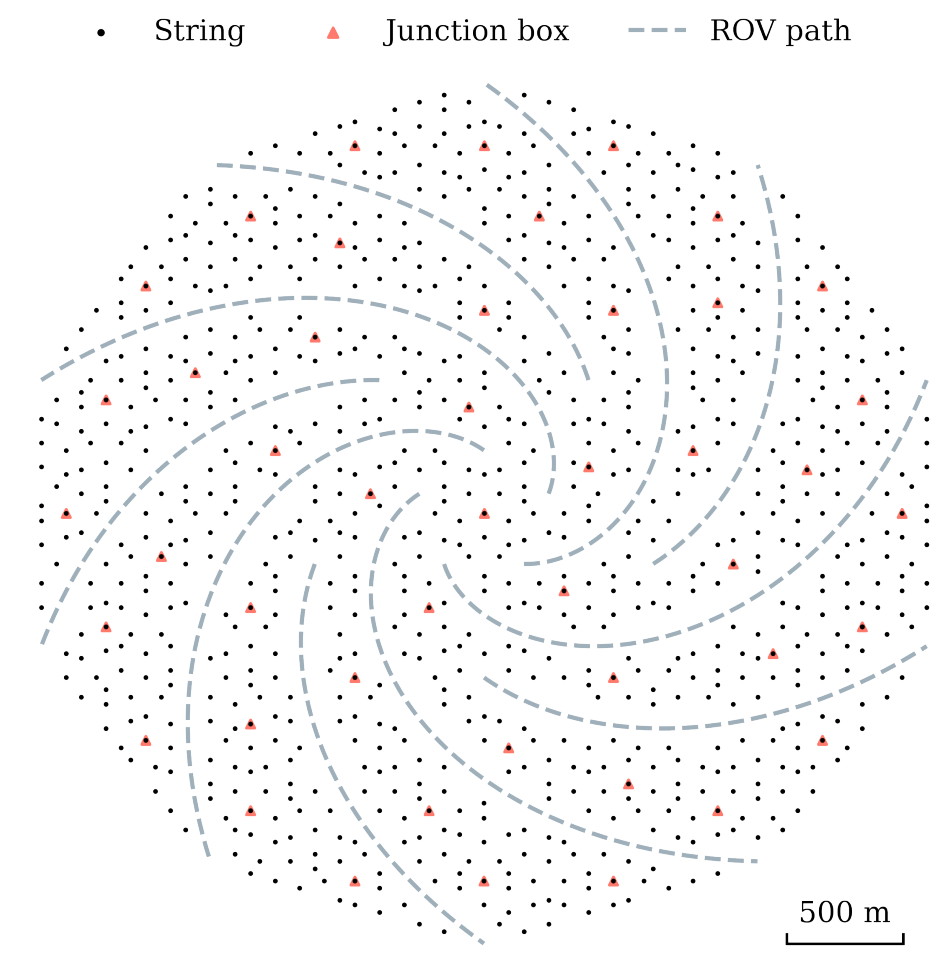} 
    \vspace{-1em} 
    \caption{Top view of TRIDENT detectors.}
    \label{fig:penrose}
\end{figure}

The detector response simulation is implemented with the Geant4 software framework \cite{AGOSTINELLI2003250,geant4}. 
Within a cylinder with a radius of 2000m, a total of  1200 vertical strings are deployed in a Penrose tiling pattern, as depicted in Figure~\ref{fig:penrose}. Each string comprises 20 hDOMs separated vertically by 30m.
During this process, the propagation and energy loss processes of particles are simulated. 
For electromagnetic cascades induced by high-energy electrons, a parameterized simulation method is employed to accelerate the simulation process, achieving a speed-up of approximately $\mathscr{O}(1000)$ times compared to traditional particle-by-particle simulations of the cascade.
For the efficient handling of Cherenkov photons, all Cherenkov photons are propagated using the OptiX ray tracing framework \cite{Opticks:2019} to utilize the acceleration of GPU.
Finaly, the detector response to Cherenkov photons is fully simulated with Geant4.

\section{Network Architecture}
In the context of neutrino telescopes, each recorded neutrino event can be intrinsically represented as a graph and can be reconstructed using GNN.
For a given event, the triggered hDOMs serve as the nodes of the graph, forming an edge-less graph.
The position (relative to the position of the initially triggered hDOM) and physics quantities of each hDOM comprises the coordinates and attributes of the corresponding node.
To establish connections between nodes,  edges are introduced such that each node is linked to its $k$ nearest neighboring nodes.  Here $k$ is a user-defined hyperparameter.
Additionally, the mean value of node attributes can serve as an indicator of the overall knowledge of a neutrino event.
Thus, a neutrino event is noted as $G=\{pos_i, x_i, e_{ij}, u\}$, where 
\begin{itemize}
    \item $pos_i$ and $x_i$ represent the location and attributes of the $i$-th hDOM, respectively.
    \item $e_{ij}$ represents the edge connecting the $i$-th and $j$-th hDOMs.
    \item $u$ is a global attribute that describes the overall characteristics of the neutrino event.
\end{itemize}

\begin{figure}[!htb]
    \centering\includegraphics[width=0.8\linewidth]{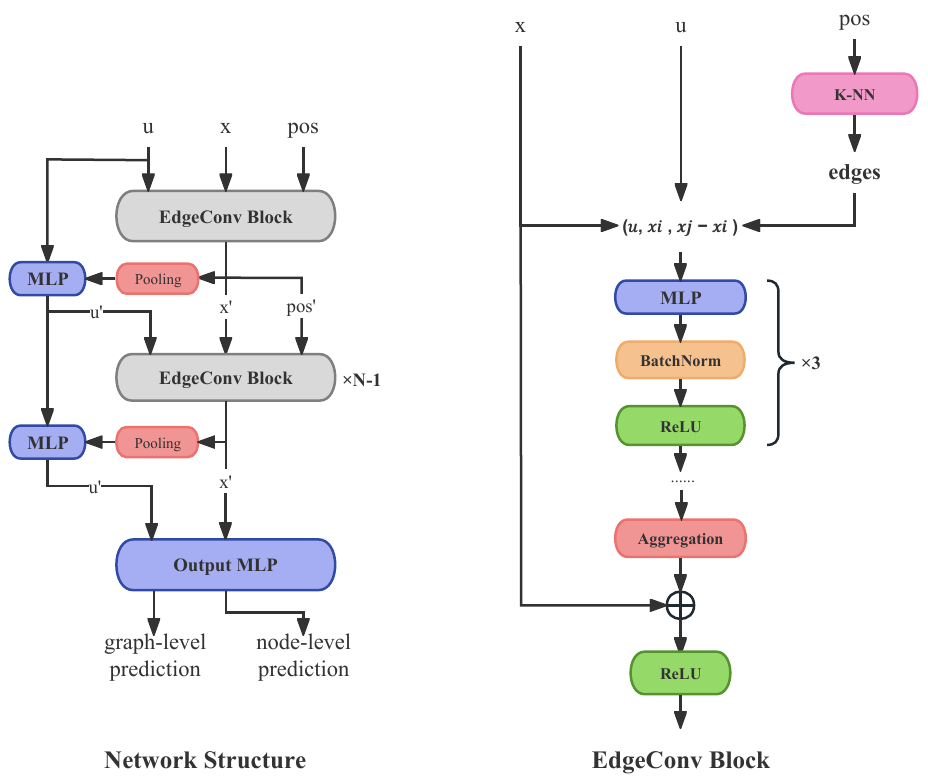} 
    \vspace{-1em} 
    \caption{Architecture of GNN used in this study is shown on the left plot. The detailed structure of EdgeConv block is illustrated on the right plot.}
    \label{fig:TridentNet}
\end{figure}

The GNN architecture utilized in this study incorporates a fundamental building block known as the EdgeConv block, as illustrated in Figure~\ref{fig:TridentNet}. This EdgeConv block is adapted from the EdgeConv block employed in ParticleNet \cite{Qu_2020}.
The EdgeConv block serves as a convolution-like operation. It commences by defining a latent vector for each edge $e_{ij}$ as: $e_{ij} = \phi_\theta(u, x_i, x_j - x_i)$. Here, $\phi_\theta$ denotes a multilayer perceptron (MLP)  with trainable parameters $\theta$.
To obtain the latent vectors for the nodes, an aggregation operation is performed based on the connected edges, which is defined as:
$x'_i = (\mathop{Max}\limits_{j=1,...k}\{e_{ij}\} + x_i \ )$.

The GNN architecture is built with several EdgeConv blocks.
In each EdgeConv block, the block updates the graph $G=\{pos_i, x_i, e_{ij}, u\}$ as follows:
\begin{align*}
    &x_i' = \text{EdgeConv}(G) \\
    &u' = \Phi_\Theta( u, \text{Global\_Average\_Pooling} ( \{x'_i\} )) 
\end{align*}
where $\Phi_\Theta$ is another MLP with parameters $\Theta$. 
By iteratively applying the EdgeConv blocks, the GNN progressively enriches the input graph with higher-level information.

The final EdgeConv block is followed by an output MLP layer for reconstructing desired physical parameters. Depending on the context, the input to this layer can be the node attributes ($x_i$) for DOM-level reconstruction or the global attributes ($u$) for event-level reconstruction.

\section{Results}
The aforementioned GNN architecture is constructed with PyTorch Geometric \cite{pytorch}\cite{pyg} and is utilized to reconstruct the direction of $\nu_e$ with 100 TeV energy and $\nu_\mu$ with energy ranges from 1TeV to 1000TeV. 
The $\nu_e$ samples are limited to a single energy level, as there is an insufficient number of samples in other energy ranges attributed to the slow speed of their simulation. 
In this section, we show the training methods and results.

\subsection{Shower-like Event Reconstruction}
For the reconstruction of $\nu_e$ events, the attribute of each node is a histogram detailing the arrival times of photons at each hDOM. 
These histograms counts the number of received photons within every 5ns time window. 
In a typical shower-like event, the majority of Cherenkov photons are received within 1000ns. Therefore, the histograms are configured to split the 1000ns interval into 200 time windows. 

The GNN model used for shower-like event reconstruction consists of of 6 EdgeConv blocks and 2 layers of output MLP and it possesses 12,289,167 trainable paramters. 
The samples are divided into training (130k samples) and validation (10k samples) sets during the training session.
The model is trained to directly predict the direction of $\nu_e$, $\vec{n}_{\nu_e}$, with MSELoss as loss function.
Subsequently, the trained models undergo testing using an additional 130k samples, yielding the results presented below.

The training result, shown in Figure~\ref{fig:shower_angular}, demonstrates the angular error between the true $\nu_e$ direction and the reconstructed direction. 
The median angular error, as is represented by the red line, is about 1.3 degrees. 
For comparison, the median angular error of 100 TeV $\nu_e$ events using the traditional likelihood method is found to be about 1.7 degrees~\cite{km3netMLE}.

\begin{figure}[H]
    \centering\includegraphics[width=0.6\linewidth]{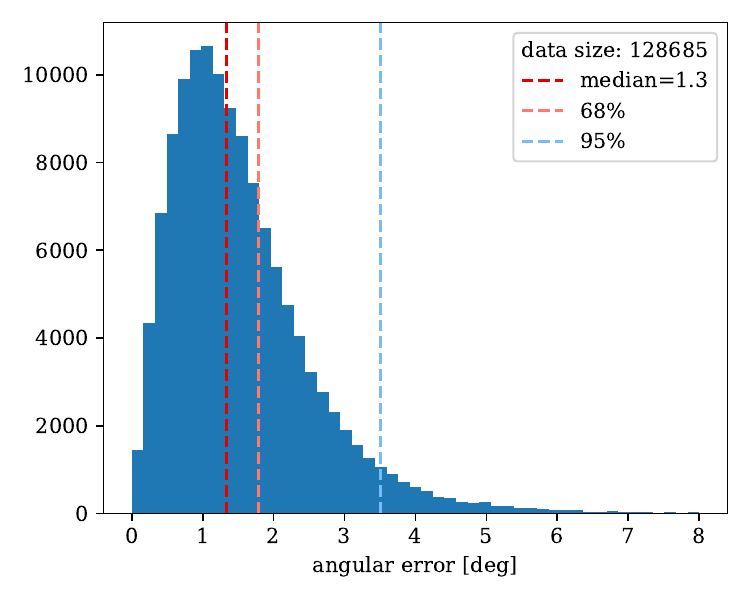} 
    \vspace{-1em} 
    \caption{The angular error for 100 TeV $\nu_e$ CC events. The red line represents the median angular error. The orange and blue lines exhibit the 68\% and 90\% quantiles. }
    \label{fig:shower_angular}
\end{figure}

\subsection{Track-like Event Reconstruction}
Muons in $\nu_\mu$ CC events leave tracks within the telescope.
The photons received by the hDOM with early arrival time are more likely to reach the hDOM with less scattering along the travelling path.
Therefore the arrival time of the photons provides useful information when reconstructing the neutrino direction and it is taken as a node attribute. To obtain the information about the distance from the hDOM to muons, the number of photons received by each hDOM is also taken as a node attribute.

\begin{figure}[!htb]
    \centering\includegraphics[width=0.3\linewidth]{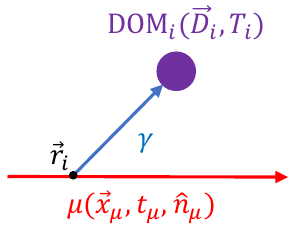} 
    \caption{Muon emits Cherenkov photon at $\vec{r}_i$ and triggers DOM$_i$.}
    \label{fig:muon_track_with_hDOM}
\end{figure}

The GNN model used for track-like event reconstruction is made of 5 EdgeConv blocks followed by 2 MLP layers (7,966,005 trainable parameters). The model is trained with MSELoss as loss function to predict the photon emission positions, $\vec{r}_i$ (as illustrated in Figure~\ref{fig:muon_track_with_hDOM}), for all triggered hDOMs. Subsequently, the direction of the muon is then reconstructed using a linear fit on the $\vec{r}_i$ positions.

In the low-energy range, graphs may have as few as 2 nodes, making it challenging to train the GNN.
To address this, all training samples must consist of more than 7 nodes. 
The dataset is partitioned into training (210k samples) and validation (70k samples) sets. 
In the evaluation phase, the trained models are tested with an additional 220k samples, each comprising more than 2 nodes, to generate the results.

\begin{figure}[H]
    \centering\includegraphics[width=0.6\linewidth]{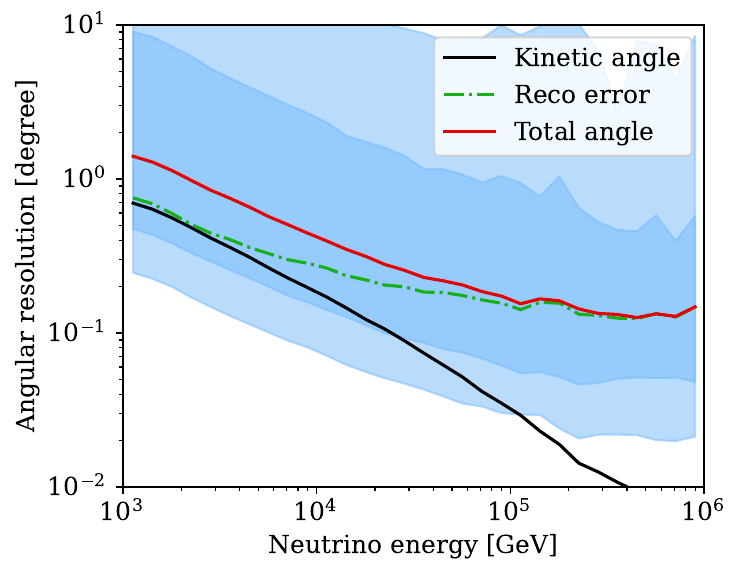} 
    \vspace{-1em} 
    \caption{
    The angular resolution of $\nu_\mu$ CC events as a function of energy. The median angle between the reconstructed track and the true direction of $\mu$ and $\nu_\mu$ is visualized by the green and red line, respectively. Color bands exhibits the 68\% and  90\% quantiles. Black line represents the median angle between direction of $\mu$ and $\nu_\mu$. }
    \label{fig:track_angular}
\end{figure}

As the result, the distribution of angular error between the true $\nu_\mu$ direction and the reconstructed direction is shown in Figure~\ref{fig:track_angular}. 
The red line represents the median angular error and the color bands exhibits the 68\% and 90\% quantiles.
The model achieves an angular resolution at the 0.1 degree level for $\nu_\mu$ events with sufficiently high energy. 
The angular resolution using the likelihood method also falls below 0.1 degree for sufficiently high energy events \cite{km3netMLE}.

\section{Summary}
In this paper, a GNN-based reconstruction method is proposed to reconstruct the direction of $\nu_e$ CC and $\nu_\mu$ CC events with high precision in TRIDENT.
For shower-like events, the median angular error achieved by this method is 1.3 degrees, which significantly outperforms the likelihood method result by 75\%. For track-like events, the median angular error reaches 0.1 degrees when the neutrino energy is sufficiently high, which gives a comparable performance with the likelihood method. 

For the next step, we plan to extend the GNN-based reconstruction method   to reconstruct both the direction and energy of neutrino events in a wide kinematic range.  We will also improve the  robustness of the method against experimental uncertainties and noises.

\section{Acknowledgements}
We thank for the support from Key Laboratory for Particle Astrophysics and Cosmology (KLPPAC-MoE) and Shanghai Key Laboratory for Particle Physics and Cosmology (SKLPPC). This work was supported by the Oceanic Interdisciplinary Program of Shanghai Jiao Tong University (project number SL2022MS020).

\bibliographystyle{ieeetr}
% \bibliography{ref}

\end{document}